\documentclass[aps,pre,amsmath,amssymb,twocolumn,showpacs,byrevtex,superscriptaddress]{revtex4}

\usepackage[final]{graphicx}
\usepackage{epstopdf} 
\begin{document}
\title{Investigation of the relation between 
local diffusivity and local inherent structures in the Kob-Andersen Lennard-Jones model.} 
\def\roma{\affiliation {
Dipartimento di Fisica and INFM Udr and CRS-SOFT: Complex Dynamics in Structured Systems, Universit\`a 
di Roma ``La Sapienza'', Piazzale Aldo Moro 2, I-00185, Roma, 
Italy}}
\def\india{\affiliation {Jawaharlal Nehru Centre for Advanced Scientific Research, Jakkur campus, Bangalore - 500064, India}}
\author{Emilia La Nave}\roma
\author{Srikanth Sastry}\india
\author{Francesco Sciortino}\roma

\begin{abstract}
We analyze one thousand independent equilibrium trajectories of a
system of 155 Lennard Jones particles to separate in a model-free
approach the role of temperature and the role of the explored
potential energy landscape basin depth in the particle dynamics. We
show that the diffusion coefficient $D$ can be estimated as a sum over 
over contributions of the sampled basins, establishing a connection
between thermodynamics and dynamics in the potential energy landscape
framework. We provide evidence that the observed non-linearity in the
relation between local diffusion and basin depth is responsible for
the peculiar dynamic behavior observed in supercooled states and
provide an interpretation for the presence of dynamic
heterogeneities.
\end{abstract}

\maketitle

\date{\today}
%
%\pacs{}
%
Supercooled liquids are characterized by dynamics which take place on
at least two well-separated time scales: a fast microscopic dynamics
(associated to atomic or molecular motions at fixed liquid structure)
and a slow structural relaxation time $\tau_{\alpha}$ (which requires
structural changes and diffusion processes)\cite{kobbook}. If one
considers the trajectory of the system in configuration space, such a
separation of time scales can be visualized as the exploration (on the
microscopic time) of a finite region of the potential energy landscape
(a basin), followed on a much slower pace by the exploration of distinct
basins \cite{pablonature}.  The potential energy landscape (PEL)
formalism introduced by Stillinger and Weber\cite{stillinger_pes}
provides a thermodynamic description which builds on such a picture.
The resulting expression for the liquid free energy requires
information on the number of basins, their local minimum energy
$e_{IS}$, and their shape.  Analysis of numerical simulations,
recently reviewed in Ref.~\cite{JSTAT}, has provided direct
quantification of the statistical properties of the landscape of
several models of glass-forming
liquids\cite{sastry01,heuer,otplungo,scot2,ivan,moreno}.  For the case
of the Kob-Andersen Lennard-Jones model\cite{kobandersen,stillweb},
the number $\Omega (e_{IS})de_{IS} \noindent$ of distinct basins of
energy depth between $e_{IS}$ and $e_{IS} + de_{IS}$ follows a
Gaussian distribution\cite{sastry01,heuer}
\begin{equation}
\Omega(e_{IS}) de_{IS}=e^{\alpha N}
\frac
{e^{-(e_{IS}-E_o)^2/2\sigma^2}}
{\sqrt{2 \pi \sigma^2}}
de_{IS}
\label{eq:Omega}
\end{equation}
\noindent
where $e^{\alpha N}$ is the total number of distinct basins for a
system of $N$ particles, $E_o$ is the energy scale of the distribution
and $\sigma^2$ is the variance; with $\sigma$, $E_o$ and $\alpha$
dependent only of the system number density.  Often, it is convenient
to define the configurational entropy $S_{conf}(e_{IS}) \equiv k_B
\log \Omega(e_{IS})$, which for the Gaussian landscape becomes a
quadratic function of $e_{IS}$
\begin{equation}
\frac {S_{conf}}{k_b}= \alpha N - \frac{(e_{IS}-E_0)^2}{2 \sigma^2}.
\label{eq:sconf}
\end{equation}
When $\Omega(e_{IS})$ is known, the system partition function can be
calculated\cite{JSTAT} (under the assumption of the
$e_{IS}$-independence of the basin anharmonicities) and predictions
can be provided for the $T$-dependence of the explored basin depth
$<e_{IS}(T)>$, the probability $P(e_{IS},T)$ of sampling a basin of
depth $e_{IS}$ at temperature $T$ and the configurational entropy
$S_{conf}(T)$.

The application of the Stillinger-Weber formalism to the analysis of
numerical simulations provides an appealing picture of the
$T$-dependent exploration of the landscape and a convenient method to
quantify the free energy of supercooled states.  Several recent
studies attempted, in different ways, to relate landscape properties
to dynamics
\cite{tom,silicaprl,angelani,cavagna,scala00,heuernew,schroder,harrowell}. The
outcome of these studies strongly support the reasonable hope that in
super-cooled states structural properties are strongly connected to
dynamic ones  and motivate
the present detailed and essentially unbiased study of the connection
between landscape properties and dynamics.
 
 In this Letter we study one-thousand independent runs of the well
 characterized Kob-Anderson model\cite{kobandersen}. We limit
 ourselves to the smallest system which can be simulated with periodic
 boundary conditions without changing the model's range of
 interaction, i.e. 155 particles at number density 1.2
 \cite{simulation}.  Due to the small system size, all quantities
 show large fluctuations, which we exploit to probe wide overlapping
 regions of $e_{IS}$ values under different $T$-conditions with the
 aim of disentangling the roles of $e_{IS}$ and $T$ in the dynamics.
 
\begin{figure}[t]
\centering
\includegraphics[width=.5\textwidth]{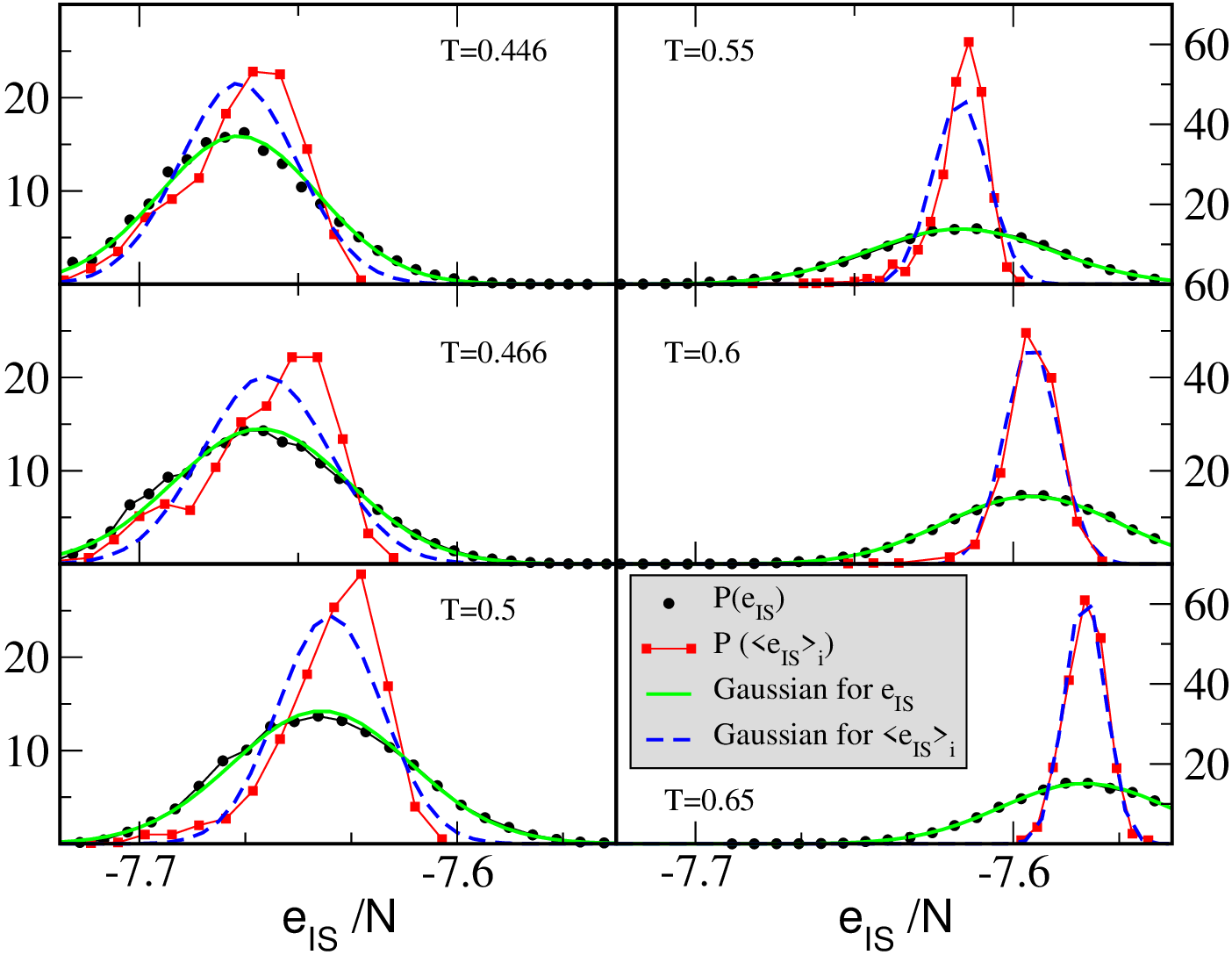}
%\vspace{-0.5cm}
\caption{Probability distribution $P(e_{IS},T)$ of
exploring an IS of depth $e_{IS}$ at temperature $T$.  Black filled
circles: equilibrium data for the N=155 system.  Green full line:
Gaussian distribution of variance $18$.  The figure also shows, with
red filled squares the distribution $P(<e_{IS}>_i,T)$ (where
$<e_{IS}>_i$ is not the instantaneous value but $e_{IS}$ averaged over
time $t_{MSD=1}$ for the trajectory $i$ and with a blue dashed line a
Gaussian distributions with the same variance as $P(<e_{IS}>_i,T)$. }
\label{fig:histo}
\end{figure}

 We simulate six different $T$ between $0.446$ and $0.65$.  In this
 supercooled $T$-window, the decay of correlation functions clearly show
the presence of two time scales and the diffusion coefficient
 $D$ varies more than two orders of magnitude. For each $T$, we
 equilibrate each independent trajectory in the canonical ensemble and
 estimate $P(e_{IS},T)$, which is found (see Fig.~\ref{fig:histo}) to
 be described rather well by a Gaussian distribution of
 $T$-independent variance $\sigma^2=18 \pm 2$, centered around a
 $T$-dependent mean $<e_{IS}>(T) \sim \sigma^2/T$, in full agreement
 with the Gaussian landscape model and with previous results
 \cite{ivan2}. The large $\sigma^2$ guarantees a significant overlap
 of the landscape region sampled at each $T$.

\begin{figure}[t]
\centering
\includegraphics[width=.44\textwidth]{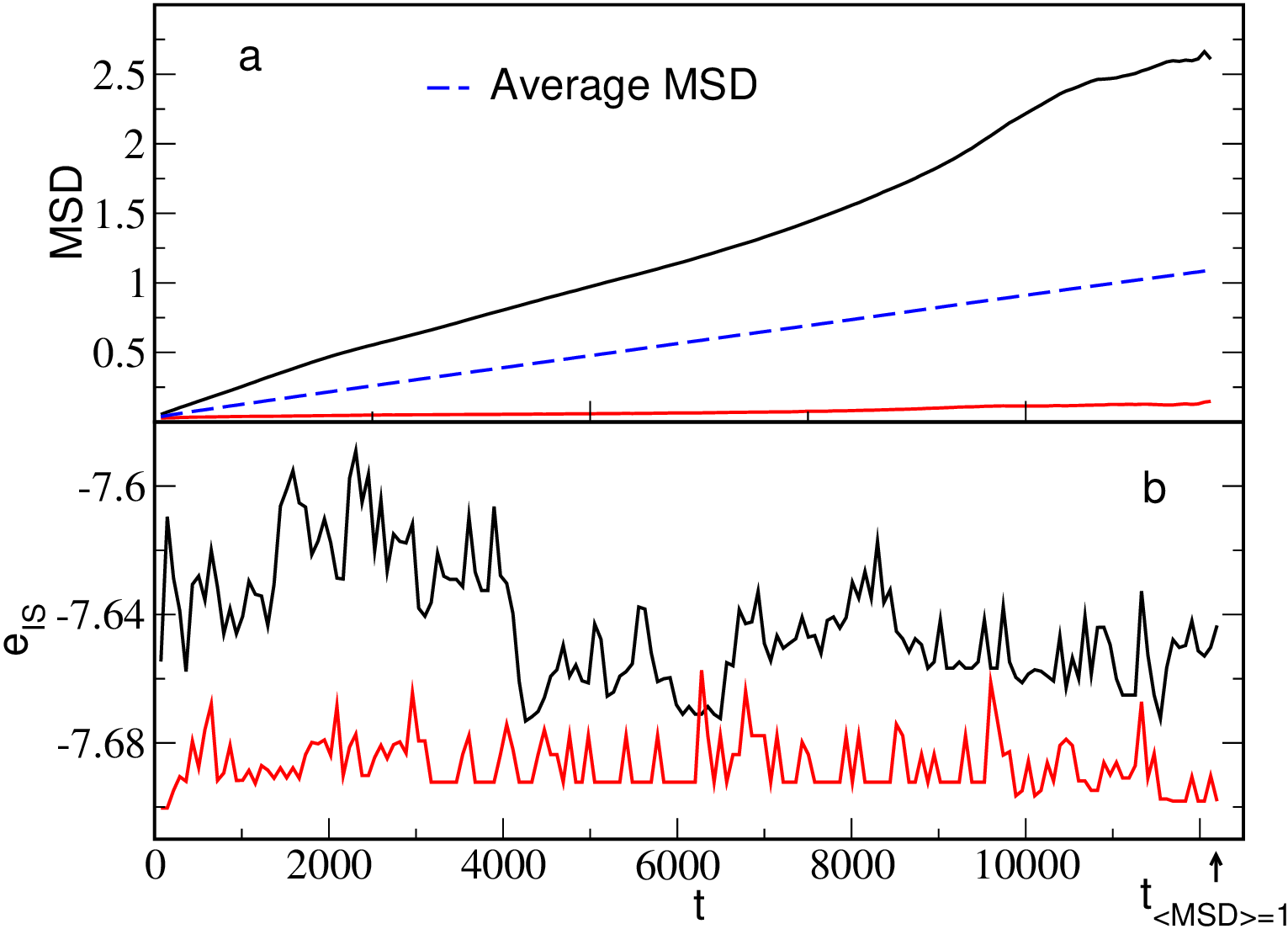}
%\vspace{-0.5cm}
\caption{a) MSD as a function of time for two different equilibrated trajectories
at the same thermodynamical state point ($T=0.446$, $\rho=1.2$). The
MSD is here averaged over all type-$A$ particles of the system as well
as over several starting times within the interval $0<t<t_{<MSD>=1}$.
The dashed line represents the MSD averaged over all trajectories and
it provides the definition of $t_{<MSD>=1}$.  b) Inherent structure
energies as a function of time for the same trajectories shown in a).}
\label{fig:eisint}
\end{figure}

Next we attempt to associate each of the independent configurations to
an $e_{IS}$ value characterizing the location of the system in the PEL
and to a $D_i$ value characterizing its local diffusion coefficient.
We further propagate each of the configurations for a time
$t_{<MSD>=1}$, defined as the time at which the type-A particles mean
square displacement (MSD), averaged over all trajectories and all $A$
particles, reaches one.  Different trajectories show very different
dynamical properties. As an example, Fig.~\ref{fig:eisint}-a
represents the MSD, as a function of $t$ for two different
trajectories at the same $T$.  Each trajectory $i$ has a rather well
defined diffusion coefficient $D_i$ which can be extracted 
from the slope of the MSD vs time. Fig.~\ref{fig:eisint}-b shows
the $t$-evolution of $e_{IS}$ for the same two
trajectories. The more mobile system is characterized by a larger and
more fluctuating $e_{IS}$. Within $t_{<MSD>=1}$, the set of $e_{IS}$
values explored is far from covering the full range of values
characteristic of that $T$. The system retains memory of the starting
configurations and this behavior is enhanced at lower $T$.  A
quantification of the location of the $i$-configuration on the PEL can
thus be provided by $<e_{IS}>_i$, defined as the value of $e_{IS}$
averaged over the time interval from zero to $t_{<MSD>=1}$.

Fig.~\ref{fig:histo} also show the probability distribution
$P(<e_{IS}>_i,T)$. While $P(e_{IS},T)$ are well approximated by a
Gaussian distribution of average $<e_{IS}(T)>$ and constant variance
(in agreement with Eq.\ref{eq:Omega}), $P(<e_{IS}>_i)$ are symmetric
and well modelled by a Gaussian only at high $T$. At low $T$
$P(<e_{IS}>_i,T)$ becomes strongly asymmetric, and its variance
increases on cooling.  This originates in the different mobility of
the system depending on the initial $e_{IS}$. Indeed, while systems
with $e_{IS}$ associated to the high energy tail of $P(e_{IS},T)$
manage to explore different basins, systems in the low energy tail
hardly lose memory of the initial configuration within $t_{<MSD>=1}$.
In this respect, the observed asymmetry confirms that there is a
strong coupling between $<e_{IS}>_i$ and $D_i$, which we now try to
quantify.
 
\begin{figure}[t]
\centering
\includegraphics[width=.5\textwidth]{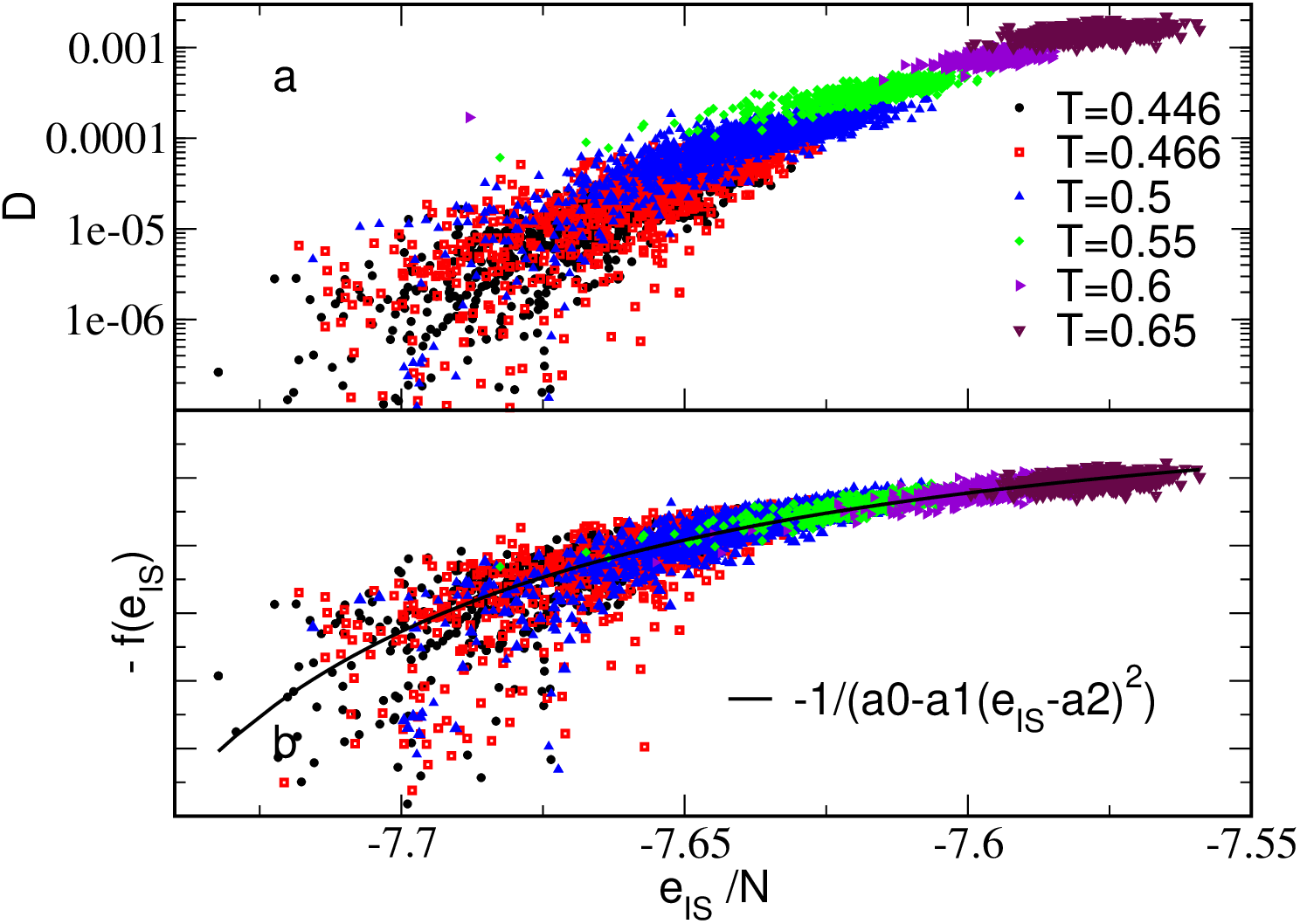}
%\vspace{-0.5cm}
\caption{Diffusivity $D_i$   as a function of
the corresponding average $<e_{IS}>_i$. Different symbols indicate the six different simulated $T$. b) $T ln(D(e_{IS},T)/D_{\infty})$ 
as a function of  $e_{IS}$. The solid line represent the fitting function $-\frac {1}{a_0-a_1 (e_{IS}-a_2)^2}$. }
\label{fig:d-all}
\end{figure}

Fig.~\ref{fig:d-all}-a shows $D_i$ as a function of $<e_{IS}>_i$ at
several $T$.  Data from different $T$ overlap in a wide window of
$e_{IS}$ values.  Data show that the local diffusivity is function of
$T$ not only via the $T$-dependence of $e_{IS}$ but also directly,
since for each chosen $e_{IS}$ value, data for higher $T$ always lay
above data at lower $T$. The fact that basins of the same depth are
sampled at different $T$ allows us to evaluate the $T$ dependence of
$D$ at fixed basin depth and sort out the relative contribution of $T$
and $e_{IS}$ to dynamics.  To distinguish between the $T$ and $e_{IS}$
dependence of $D$ we hypothesize that the law relating these two
quantities has the form:
\begin{equation}
D(e_{IS},T)=D_{\infty} e^{-\frac{f(e_{IS})}{T}},
\label{eq:deq}
\end{equation}
a form compatible both with the idea of activated dynamics as well as
with the Adam-Gibbs (AG) hypothesis $ln(D/D\infty) = \frac{-B}{T
S_{conf}}$ (in which case $1/f(e_{IS}) = S_{conf}(e_{IS})/B$), with
$B$ being a coefficient related to the minimum size of a cooperatively
rearranging region.

According to Eq.~\ref{eq:deq}, a plot of $T
ln(D(e_{IS},T)/D_{\infty})$ should produce a collapse of all data,
independently of $T$, onto a master curve which can be identified as
$-f(e_{IS})$.  Fig.~\ref{fig:d-all}-b confirms that indeed such a
procedure generates an impressive scaling of the data.  We note that,
in this procedure, we use the value of $D_{\infty}=0.155$ obtain
independently from high $T$ simulations.  If the resulting $f(e_{IS})$
is interpreted according to activated models, one has to conclude that
the height of the barriers increases significantly on cooling,
diverging as the limiting $e_{IS}$ value $e_K=-7.81$ is approached, a
result which appears to contradict, in the details, the analysis
performed previously by Doliwa and Heuer \cite{heuernew}.  If
$f(e_{IS})$ is interpreted according to the AG equation, then
$1/f(e_{IS})$ has to be proportional to $S_{conf}(e_{IS})$, a quantity
which has been previously calculated for this model using
thermodynamic integration techniques \cite{skt99,sastry01,depablo}.

\begin{figure}[t]
\centering
\includegraphics[width=.48\textwidth]{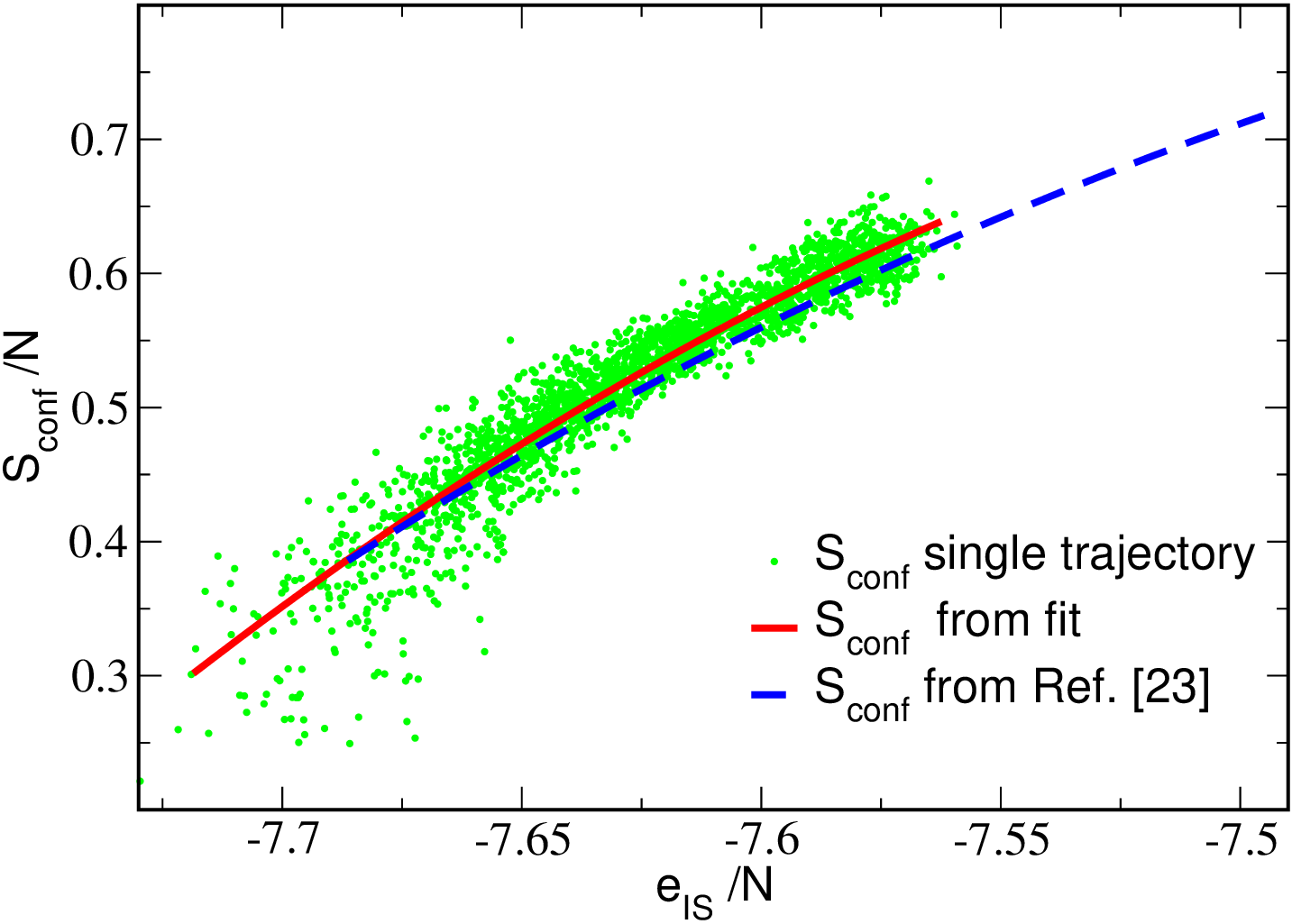}
%\vspace{-0.5cm}
\caption{Comparison between $S_{conf}$ evaluated from Eq. ~\ref{eq:sconf} and ~\ref{parameter} (solid line), and $S_{conf}$ from  \cite{skt99} (dashed line). The symbols 
are the values obtained for each trajectory $i$. }
\label{fig:sconf}
\end{figure}

The test between the two different estimates of
$S_{conf}$ can be  made even more stringent,
if one realizes that, by fitting  $f(e_{IS})$ according to
the relation  $f(e_{IS})=\frac {1}{a_0-a_1 (e_{IS}-a_2)^2}$
the fitting coefficient $a_0$, $a_1$ and $a_2$ 
can be associated, comparing with Eq.~\ref{eq:sconf}, to

\begin{eqnarray}
a_0= (\alpha N)/B  \nonumber \\
a_1=1/(2 \sigma^2 B)  \nonumber \\ 
a_2=E_0
\label{parameter}
\end{eqnarray}
Interestingly enough, since $\sigma^2$ is known independently from the
(T-independent) width of $P(e_{IS},T)$, $a_1$ provides $B$, $a_0$
$\alpha$ and $a_2$ provides $E_0$ (yielding respectively $B=1.85$,$
\alpha=0.78$,$E_0=-1140$).  Thus, an estimate of $S_{conf}$ in
absolute units is provided by the fit parameters.  The comparison
between $S_{conf}$ obtained via thermodynamic integration (from
Ref.\cite{skt99}) and $S_{conf}$ obtained here from $f(e_{IS})$ is
shown in Fig.~\ref{fig:sconf}. The agreement between the two
independent estimates of $S_{conf}$ confirms the previously observed
possibility of describing the dynamics in this model according to the
AG equation\cite{sastry01} and to the interpretation of the landscape
basins as states\cite{wolynes}.

The combined use of the PEL thermodynamic formalism and of the
relation between thermodynamic and dynamics provided by
Eq.~\ref{eq:deq} makes it possible to evaluate the $T$-dependence of
$D$ for the present model, --- independently from the interpretation
in terms of activated processes with $e_{IS}$ dependent barrier or in
term of AG --- according to

\begin{equation}
D(T) = \int D(e_{IS},T) P(e_{IS},T) de_{IS}.
\label{eq:d}
\end{equation}

\begin{figure}[t]
\centering
\includegraphics[width=.42\textwidth]{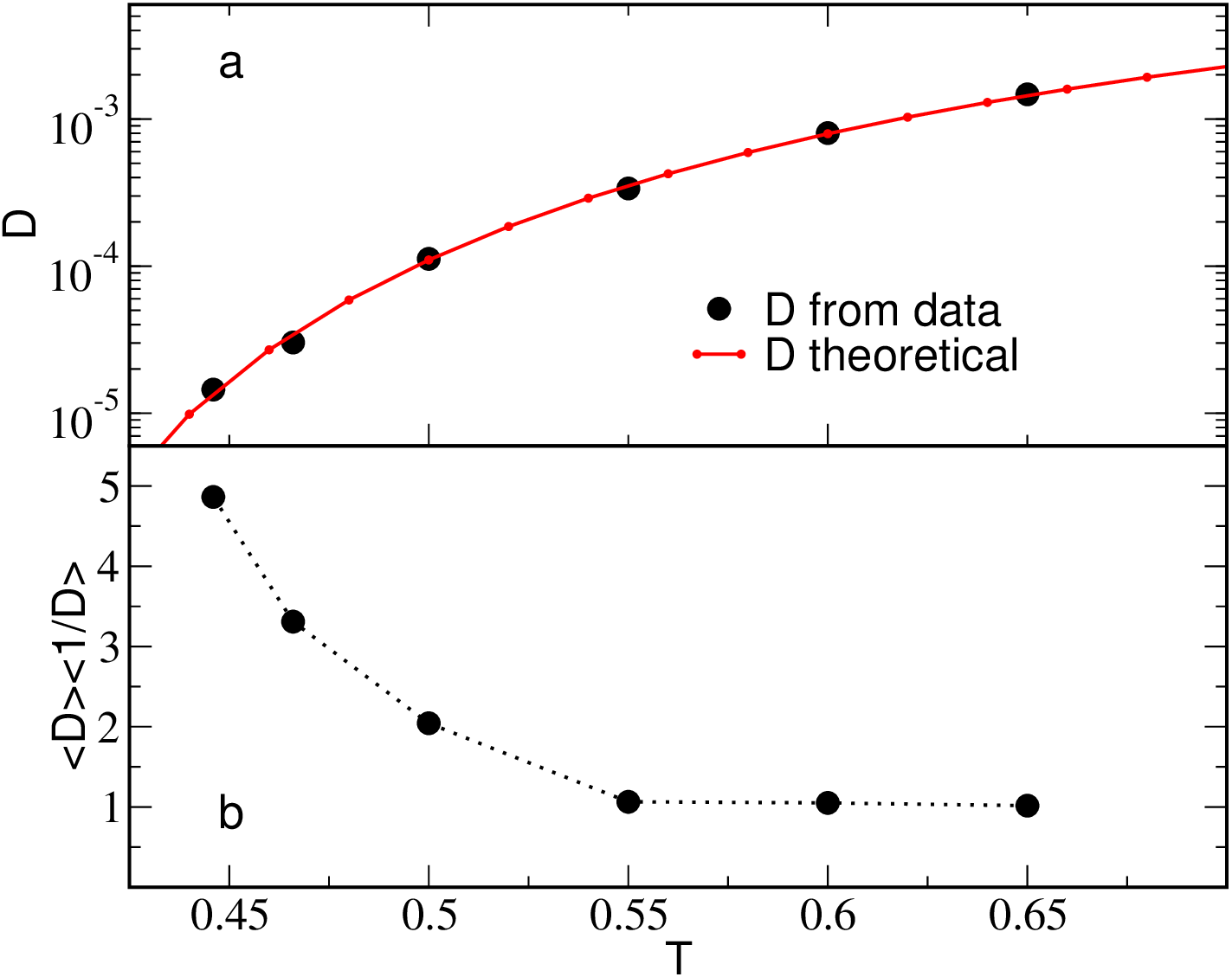}
%\vspace{-0.5cm}
\caption{a) Diffusivity constant as from data compared with the diffusivity constant evaluated from Eq. \ref{eq:d}.
b) Product $<D><1/D>$ form data as a function of $T$. The dashed line is a guide for the eye.}
\label{fig:D-ricos}
\end{figure}

Fig.~\ref{fig:D-ricos} shows $D$ evaluated directly from the MSD long
time limit and from Eq.~\ref{eq:d}. The agreement between the two sets
of independent data strongly support the possibility of representing
$D$ as a sum over the local diffusivity of different parts of the
landscape. This representation is very intriguing and offers an
alternative way, within the landscape framework, to look at dynamic
heterogeneities, which have been widely studied theoretically, in
simulations and experiments in recent years
\cite{stillingerandh,tarjus,harro,scg,onuki,mik,gc,cicerone,Chang,weeks,kegel,ediger,bordat,szamel}. Indeed,
the nonlinear relation between $D$ and $e_{IS}$ (Eq.~\ref{eq:deq})
implies that at low $T$, in equilibrium, the system can be represented
by coexisting regions in space with very different local mobility. As
seen in Fig.~\ref{fig:d-all}, at the lowest investigated $T$, the
$D_i$ values cover about three different orders of magnitude. It is
important to stress that the inhomogeneities in the dynamics are, in
the present case, strongly associated to structural properties and in
particular to the depth of the lS locally sampled.

This extremely wide distribution of $D_i$ values suggests also a
possible explanation of the $T$-dependence of the product
$D\tau_{\alpha}$ upon supercooling, analyzed with reference to
heterogeneous dynamics in, {\it e. g.}
\cite{stillingerandh,tarjus,onuki,mik,gc}, along the lines proposed in
Ref.~\cite{stillingerandh}. Indeed, when a distribution is wide, the
mean and the inverse of the average of its inverse can be rather
different, since they weigh different parts of the distribution.  To
evaluate this effect in the present case we compare in
Fig.~\ref{fig:D-ricos} the $T$ dependence of the product $<D><1/D>$.
On cooling, the product starts to deviate from one, consistently with
the observations for the model liquid studied here that the product $D
\eta/T$ \cite{bordat} and $D \tau_{\alpha}$\cite{szamel} starts to increase on cooling.

In summary, by considering large numbers of realizations of the
trajectory of a small system of particles, we have shown that the
dependence of the diffusion coefficients for each such trajectory on
the depth of the potential energy minima explored can be extracted in
a model-free manner, and that the average diffusion coefficient for
the system can be estimated as a sum over contributions of the
sampled basins, establishing a connection between thermodynamics and
dynamics in the potential energy landscape framework. We provide
evidence that the observed non-linearity in the relation between local
diffusion and basin depth is responsible for the peculiar dynamic
behavior observed in supercooled states and provide an interpretation
within the energy landscape picture for the presence of dynamic
heterogeneities.

We acknowledge support from MIUR-FIRM. S. Sastry
thanks the University of Roma for hospitality. 

%Correspondence should be addressed to francesco.sciortino@phys.uniroma1.it
%
%%%%%%%%%%%%
% REFERENCES
%%%%%%%%%%%%
%
%\end{document}

%
%%%%%%%%%%
% FIGURES
%%%%%%%%%

%

\begin{references}
\bibitem{simulation}
We simulate a system of 155 particles in a volume $V=(5.04938)^3$
corresponding to a reduced density of 1.2 .  The system is a binary
mixture of particles A ($80$ \%) and B ($20$ \%).  The interaction
potential (including the cut-off values) coincides with the one
introduced in Ref.~\cite{kobandersen}.  Units of energy and length are
defined by the $\sigma$ and $\epsilon$ parameters of the $A-A$ Lennard
Jones interaction potential and the unit of mass the mass of the A particles.
Temperatures are expressed in units of $\epsilon/k_B$.  Determination of the IS
energy has been performed using a standard conjugate-gradient
minimization algorithm.
\end{references}
\end{document}